\documentstyle[aps,epsf,eqsecnum,floats,preprint,tighten]{revtex}
\begin{document}
%
\def\pr#1#2#3{ {\sl Phys. Rev.\/} {\bf#1}, #2 (#3)}
\def\prl#1#2#3{ {\sl Phys. Rev. Lett.\/} {\bf#1}, #2 (#3)}
\def\np#1#2#3{ {\sl Nucl. Phys.\/} {\bf B#1}, #2 (#3)}
\def\cmp#1#2#3{ {\sl Comm. Math. Phys.\/} {\bf#1}, #2 (#3)}
\def\pl#1#2#3{ {\sl Phys. Lett.\/} {\bf#1}, #2 (#3)}
\def\apj#1#2#3{ {\sl Ap. J.\/} {\bf#1}, #2 (#3)}
\def\aop#1#2#3{ {\sl Ann. Phy.\/} {\bf#1}, #2 (#3)}
\def\nc#1#2#3{ {\sl Nuo. Cim.\/} {\bf#1}, #2 (#3)}
\newcommand{\beq}{\begin{equation}}
\newcommand{\eeq}{\end{equation}}
\newcommand{\bea}{\begin{eqnarray}}
\newcommand{\eea}{\end{eqnarray}}  
\newcommand{\aprime}[1]{#1^\prime}
\newcommand{\daprime}[1]{#1^{\prime\prime}}
\newcommand{\aoy}{\frac{\alpha}{y}}
\newcommand{\poy}{\frac{\varphi}{y}}
\newcommand{\noy}{\frac{\nu}{y}}
\newcommand{\dap}{\delta a^\prime}
\newcommand{\dapp}{\delta a^{\prime\prime}}
\newcommand{\dnp}{\delta n^\prime}
\newcommand{\dnpp}{\delta n^{\prime\prime}}
\newcommand{\dpp}{\delta \phi^\prime}
\newcommand{\dbp}{\delta b^\prime}
\newcommand{\Vdb}{{Ve^{\bar b \phi_0}(\delta b+\bar b\delta\phi)}}
\newcommand{\ie}{{\it i.e.}}
\newcommand{\eg}{{\it e.g.}}
\newcommand{\arcsinh}{{\rm arcsinh}}
\preprint{\vbox{\hbox{UCSD-PTH-00-32} 
    \hbox{IASSNS-HEP-00/84} \hbox{MIT-CTP-3058}}}
\title{On a Covariant Determination of Mass Scales in Warped Backgrounds}
\author{Benjam\'\i{}n Grinstein\thanks{e-mail addresses: 
   {\tt bgrinstein@ucsd.edu, nolte@ias.edu, skiba@mit.edu}}, 
Detlef R. Nolte$^\dagger$, and
Witold Skiba$^\ddagger$}
\address{$^*$Department of Physics,  University of California at San Diego,
         La Jolla, CA 92093 \\
         $^\dagger$Institute for Advanced Study, Princeton, NJ 08540 \\
         $^\ddagger$Center for Theoretical Physics, Massachusetts
         Institute of Technology, Cambridge, MA 02139 }
\date{December 7, 2000}

\maketitle
\begin{abstract}
We propose a method of determining masses in brane scenarios which is
independent of coordinate transformations. We apply our method to
the scenario of Randall and Sundrum (RS) with two branes, which
provides a solution to the hierarchy problem. The core of our proposal is the 
use of covariant equations and expressing all coordinate quantities in
terms of invariant distances. In the RS model we find that massive brane
fields propagate proper distances inversely proportional to masses that
are not exponentially suppressed. The hierarchy between the gravitational
and weak interactions is nevertheless preserved on the visible brane
due to suppression of gravitational interactions on that brane. 
The towers of Kaluza-Klein states for bulk fields are observed to have
different spacings on different branes when all masses are measured in
units of the fundamental scale. Ratios of masses on each brane are
the same in our covariant and the standard interpretations.  
Since masses of brane
fields are not exponentiated, the fundamental scale of higher-dimensional
gravity must be of the order of the weak scale.

\end{abstract}

\section{Introduction}
\label{sec:intro}
Brane scenarios with nontrivial gravity backgrounds have recently attracted
a lot of attention. We will focus on the proposal of Randall and 
Sundrum~\cite{RS} with a single extra dimension and two branes, but 
our arguments can be applied to any model with warped geometry.
The standard way of calculating masses of brane and bulk fields,
which we will review shortly, is to absorb metric factors by a field
redefinition such that the kinetic terms are canonically normalized.

We consider observers that live on branes and assume that the brane
metric is the induced metric
\begin{equation}
 g_{\rm brane}=G|_{y=y_{\rm brane}},
\end{equation} 
where $G$ is the metric on the full extra dimensional space. The case
of observers that live in the bulk will be discussed in a forthcoming
paper. 
We will use $x^{\mu}$ to denote four dimensional coordinates, and
$y$ the fifth dimension. Brane observers, by definition,
measure distances along the brane using the induced metric.
The two mass parameters one needs to determine in any brane scenario
are the masses of brane fields that do not propagate in extra dimensions
and the scale of gravitational interactions.

We propose to determine all masses by using covariant equations and
then expressing all coordinate distances in terms of the proper
distances. To determine the masses of brane fields we use the two-point
function and measure the rate of exponential falloff in the Euclidean domain.
For gravitational interactions, we use linearized perturbations around the
background to determine the acceleration of test particles infalling
into point-like sources. 

Using this covariant procedure we find that brane masses are
independent of the warp factor. That is brane fields with Lagrangian
mass $M$ are measured by brane observers to have mass $M$ no matter
where the brane is located. Newton's constant observed on different
branes is proportional to the warp factor on the given brane. This
follows the intuitive picture of suppression of gravitational
interaction due to the graviton wavefunction, which coincides with the
warp factor. Thus, the hierarchy on the visible brane is realized the
same way it is realized in the conventional interpretation. That the
fundamental scale of higher dimensional gravity is exponentially
smaller than that of four dimensional gravity was presented as one of
two alternative interpretations in \cite{RS}.  The covariant
interpretation indicates that this is the proper interpretation, and
follows if one demands that the same units be used to measure
quantities on both branes and in the bulk.

The covariant interpretation allows for straightforward analysis of
relative scales on the branes. Consider the following toy model. In
accord with the proposed solution to the hierarchy problem include
actions on the visible and hidden branes with dimensionful parameters
given in terms of one mass scale only, $M_5$, the scale of the
underlying five-dimensional gravity. For simplicity both actions are
identical and contain a scalar field of mass parameter $M_5$ and a
gluon field with the same dimensionless coupling $g(M_5)$. There is
also a bulk scalar field of mass $M_5$. What is the spectrum of this
theory? The covariant approach gives the answer immediately. Each
brane-scalar field gives a spinless particle of mass $M_5$ in its own
brane. The gluons give glueballs, with the same spectrum on both
branes, at a mass scale $\Lambda_{\rm
QCD}=M_5\exp\left(-\int_\infty^{g(M_5)}dg/\beta(g)\right)$. Gravity
and the bulk-scalar field finally give a difference between the two
branes. The effective strength of gravity in the hidden brane is the
same as in the underlying five dimensional gravity, but in the visible
brane gravity is exponentially weaker. Similarly, to an observer on
the hidden brane there is a tower of states with exponentially
suppressed masses while an observer on the visible brane sees these
excitations as unsuppressed. The same units have been used to describe
the spectrum on both visible and hidden branes which is important
since meaningful comparisons of masses between the two branes are
possible through gedanken experiments.

Before we present any further details let us briefly review the RS model.
Consider a $Z_2$ 5-dimensional orbifold with 3-branes at the fixed points.
Furthermore, assume there is a negative cosmological constant $\Lambda$ and
tension on the branes, $V_{{\rm hid}}$ and $V_{{\rm vis}}$. Then the  metric
\beq
\label{eq:RSmetric}
ds^2=G_{MN}dx^Mdx^N=a^2(y)\eta_{\mu\nu}dx^\mu dx^\nu - dy^2 
\eeq 
solves Einstein's equations provided \beq V_{hid}=-V_{vis}=24M^3 k
\qquad \Lambda=-24M^3k^2 \qquad a(y)= e^{-k|y|} \eeq where $M$ is the
fundamental 5-dimensional gravitational mass scale. The brane with
negative tension, $V_{vis}$, contains the visible universe and is
located at $y=y_c$ while the hidden brane, with tension $V_{hid}$, is
at $y=0$. Now consider a scalar field $\varphi$ on the visible
brane. Its action is of the form 
\beq 
\label{eq:Svis0}
S_{\rm vis}=\int d^4x \;\sqrt{-g}(
{\scriptstyle\frac12} g^{\mu\nu}\partial_\mu\varphi\partial_\nu\varphi 
-\varphi^4V(\varphi/M)).
\eeq
The only available mass scale is $M$, and it is assumed that the function
$V(z)$ does not contain any anomalously large or small numerical
constants. Using the four dimensional part of the metric
Eq.~(\ref{eq:RSmetric}) the action is
\beq 
\label{eq:Svis}
S_{\rm vis}=\int d^4x\; a^4(y_c)(
{\scriptstyle\frac12} a^{-2}(y_c)\eta^{\mu\nu}\partial_\mu\varphi\partial_\nu\varphi 
-\varphi^4V(\varphi/M)).
\eeq
The kinetic energy is not canonically normalized. Rescaling the fields
$\varphi\to\varphi/a(y_c)$
one obtains
\beq
\label{eq:Svis-rescaled}
S_{\rm vis}=\int d^4x\; (
{\scriptstyle\frac12} \eta^{\mu\nu}\partial_\mu\varphi\partial_\nu\varphi 
-\varphi^4V(\varphi/a(y_c)M)).
\eeq
The only scale that appears in this action is $a(y_c)M$, which is
exponentially small compared to $M$ provided $ky_c>1$.

To complete the argument that a hierarchy has been generated Randall
and Sundrum verify that the effective 4-dimensional Planck mass
$M_{\rm Pl}$ is not exponentially small compared to $M$. To this effect
the four dimensional zero mode of the metric $\bar h_{\mu\nu}$ is
introduced through
\beq
\label{eq:RSmetrich}
ds^2=a^2(y)(\eta_{\mu\nu}+\bar h_{\mu\nu} )dx^\mu dx^\nu - dy^2.
\eeq 
The four dimensional metric is $\bar g_{\mu\nu}=\eta_{\mu\nu}+\bar
h_{\mu\nu}$. The curvature term in the 5-dimensional action
contains the four dimensional curvature term, so the effective four
dimensional theory, at distances large compared to the size of the
fifth dimension is
\beq
\label{eq:Seff}
S_{\rm eff}= \int d^4 x \int_0^{y_c} dy \;2M^3 a^2(y)\sqrt{-\bar g}\bar R
\eeq
Since $\bar g$ is $y$ independent the integral over $y$ is trivial and
gives
\beq
M^2_{\rm Pl}=\frac{M^3}k[1-a^2(y_c)].
\eeq
Since $k$ is naturally of order $M$, then also  $M_{\rm Pl}\sim M$. 

It is apparent that the above interpretation of masses for brane fields
depends on the choice of coordinates. For example, one can transform
coordinates globally, not just on the brane, such that the masses of
brane fields are unchanged on both branes. Take
\beq
x^{\prime\mu}=f(y)x^\mu\qquad y'=y,
\eeq
and choose the  function $f(y)$ to satisfy
\beq
f(y)=\cases{1& \hbox{if $y<y_c/2-\epsilon$}\cr 
         a(y_c)& \hbox{if $y>y_c/2+\epsilon$}\cr}
\eeq
and to interpolate smoothly between these values for
$|y-y_c/2|\le\epsilon$.  Then, in the new coordinate system
the actions $S_{\rm vis}$ as well as $S_{\rm hid }$ are canonically normalized.
Therefore no mass rescaling takes place on either brane.
Notice that the transformation is consistent with the orbifold
construction (it leaves the branes fixed).

In the new coordinates the 5-dimensional metric fails to be diagonal
only in the region $|y-y_c/2|<\epsilon$. As long as we are not interested
in gravity, the brane observers would not be aware that the metric
is quite complicated in the bulk.

The plan of the paper is as follows. First, in Sec.~\ref{sec:rescaling}
we will explain why the masses in brane actions are not exponentially
suppressed in {\it any} coordinate system. Next, in Sec.~\ref{sec:Planck}
we show that  Newton's constant depends on the position of the brane.
Roughly speaking,  Newton's constant is proportional to the warp factor
at the position of the brane. In Sec.~\ref{sec:KK} we consider bulk
scalar fields. We show that the towers of Kaluza-Klein states 
are measured to have different masses on the two branes. It may be surprising that
the same bulk state is observed by different observers to have different
mass, but this is nothing else but the gravitational
red shift. For completeness, in the Appendix we derive the Green
function of massless and massive bulk scalar fields.

None of the calculations in the paper are original, with the
possible exception of the simple derivation in the Appendix.
In Sec.~\ref{sec:Planck} we rely on the results of
Refs.~\cite{GarrigaTanaka,TanakaMontes}.
The derivation in the Appendix is a simple modification of a result from
Ref.~\cite{GKR} to the case with two branes.

\section{Rescaling and masses on the branes}
\label{sec:rescaling}
Consider a flat four dimensional space with metric 
\beq
\label{eq:trivialmetric}
ds^2=\eta_{\mu\nu}dx^\mu dx^\nu.  
\eeq 
We would like to study particle propagation in this theory, and to
compare results computed with this metric and those obtained in a
different coordinate system $x^{\hat \mu}=\delta^{\hat \mu}_\mu\lambda
x^\mu$, so that 
\beq
\label{eq:dslambda}
ds^2=\lambda^2\eta_{\hat\mu\hat\nu}d{x}^{\hat\mu} d{x}^{\hat\nu}.
\eeq

A free scalar field $\phi$ has action
\bea
S &=& \int d^4x\;({\textstyle\frac12}\eta^{\mu\nu}\partial_\mu\phi
\partial_\nu\phi-{\textstyle\frac12}m^2\phi^2)\\
&=& \int d^4\hat{x}\;\lambda^4({\textstyle\frac12}\lambda^{-2}
\eta^{\hat\mu\hat\nu}\partial_{\hat\mu}\phi
\partial_{\hat\nu}\phi-{\textstyle\frac12}m^2\phi^2)
\eea
Rescaling the field, $\phi\to\lambda^{-1}\phi$ the kinetic energy is
canonically normalized,
\beq
\label{eq:Slambda}
S=\int d^4\hat{x}\;({\textstyle\frac12}
\eta^{\hat\mu\hat\nu}\partial_{\hat\mu}\phi
\partial_{\hat\nu}\phi-{\textstyle\frac12}(\lambda m)^2\phi^2).
\eeq
It would seem the field describes a particle of mass $\lambda m$.
However, consider the two-point function obtained by
solving the covariant equation
\begin{equation}
  (\Box +m^2)G(x,x')=\frac{\delta^4(x-x')}{\sqrt{g}},
\end{equation}
where $\Box$ is the scalar Laplacian in the background of
(\ref{eq:dslambda}).  The corresponding Feynman propagator
$\Delta^{(4)}(\hat x,\hat x')$ is a function of $\hat\sigma\equiv
\eta_{\hat\mu\hat\nu}(x- x')^{\hat\mu}(x- x')^{\hat\nu}$ only,
\beq
\Delta^{(4)}(\hat x,\hat x') =\frac{i}{8\pi}
\sqrt{\frac{(\lambda m)^2}{-\hat\sigma+i\epsilon}}
H^{(2)}_1(\sqrt{(\lambda m)^2(\hat\sigma-i\epsilon)}),
\eeq
where $H^{(2)}_1$ is a Hankel function. We have included the small
imaginary part, $i\epsilon$, to remind us that $\Delta^{(4)}$ is
really the boundary value of a function that is analytic in the lower
$\hat\sigma$ complex plane. One can extract the mass by looking for
the exponential fall-off in the Euclidean domain, $\Delta^{(4)}\sim
\exp(-\lambda m \sqrt{-\hat\sigma})$. In terms of the physical
Euclidean separation $d_E=\sqrt{-\sigma}=\lambda\sqrt{-\hat\sigma}$,
the exponential fall-off is $\exp(-md_E)$. The Green function
falls-off exponentially over a physical length scale $1/m$, not
$1/\lambda m$. As we have stressed already, it is crucial that we 
do use the background metric to convert to physical, coordinate 
independent, distances.

The situation is entirely analogous in the RS model.  Identifying
$\lambda=a(y_c)=\exp(-ky_c)$ the action on the visible brane,
Eq.~(\ref{eq:Svis}), has all masses rescaled $M\to\lambda M$ as in
the previous section and describes propagation in a 
background $ds^2=a^2(y_c)\eta_{\mu\nu}dx^\mu dx^\nu$. One can infer
the physical mass by studying propagation and looking for the
exponential fall-off in physical separation concluding that the mass
is order $M$ rather than $a(y_c)M$, or more simply by rescaling $x\to
\exp(ky_c)x$, so
\beq
\label{eq:RSmetricrescaled}
ds^2=e^{2k(y_c-|y|)}\eta_{\mu\nu}dx^\mu dx^\nu-dy^2.
\eeq

We note that in more general situations rescaling of the field is not
sufficient to bring the action on the visible brane into standard
form, while a change of coordinates does. There exist background
solutions\cite{Binetruy:2000ut} where the time and space components of
the metric are not identical. One can have
\beq
ds^2=n^2(y,t)dt^2-a^2(y,t)\delta_{ij}dx^idx^j - b^2(y,t)dy^2.
\eeq
with ${a^2}\ne{n^2}$ being functions of $y$ and $t$.  For static
backgrounds, see for instance Ref.~\cite{us}, the brane is flat and
the effective action for a brane scalar field is
\beq 
\label{eq:Seffwmatter}
S_{\rm vis}=\int d^4x a(y_c)^3n(y_c)\left[
\frac12 \left(\frac{1}{n^2(y_c)}\partial_t\varphi\partial_t\varphi 
-\frac{\delta^{ij}}{a^2(y_c)}\partial_i\varphi\partial_j\varphi\right) 
-\varphi^4V(\varphi/M)\right].
\eeq
It is apparent that no  field rescaling will bring the action into
the canonical form. However, a coordinate transformation can be performed
to this effect. More generally, if the 3-brane is flat there is a
coordinate system for which the metric is~(\ref{eq:trivialmetric}).

\section{Planck Scale for Brane Observers}
\label{sec:Planck}
We want to find out the effect of point masses placed on
branes on test particles also placed on the branes.
Assuming that the sources do not significantly affect the background
one can perform a computation of linearized perturbations of the metric.
For the RS scenario, such a calculation has been presented
in Refs.~\cite{GarrigaTanaka,TanakaMontes}.
Compare also Ref.~\cite{GKR} for the linearized gravity calculation
in the one-brane RS model~\cite{RS2}. Ref.~\cite{GarrigaTanaka}
restricts the calculation to metric fluctuations of spin two neglecting
the scalar excitations. Ref.~\cite{TanakaMontes} shows that the scalar
modes effectively do not contribute to long distance forces
provided that the radius is stabilized~\cite{GW:radius,DFGK}.

We are only interested in the long distance gravitational interactions,
we can therefore neglect all higher Kaluza-Klein states as well as
massive fields with spins lower than two. It is however crucial
that there are no such massless fields coming from the dimensional
reduction of the five-dimensional metric tensor. If there were such
massless fields they might provide dominant contribution to the 
gravitational interactions and actually destroy the hierarchy.
For example, the radion coupling
on the visible brane are enhanced compared to the spin-two excitations
\cite{Rubakov} and the radion, if massless, would provide the dominant
force at long distance. Of course, if this was the case the model
would be ruled out because scalar gravity does not bend light.

Our strategy is as follows. Newton's constant (in four dimensions)
$G_N$ is defined, in the non-relativistic limit, as the
proportionality constant that gives the acceleration of a particle in
the gravitational field of a point  mass $m$,
\beq 
\label{eq:newtonslaw}
\frac{d^2\vec x}{dt^2}=-G_N m\frac{\vec x}{|\vec x|^3}.  
\eeq 
We will compute this equation taking the non-relativistic
approximation of the geodesic equation for the source and a particle both
constrained to one of the branes.

In terms of the metric perturbations
\beq
ds^2=(a^2(y) \eta_{\mu \nu}+{h}_{\mu \nu}) dx^\mu dx^\nu - dy^2
\eeq
the authors of Refs.~\cite{GarrigaTanaka,TanakaMontes}
find that the long distance interactions are governed by
\beq 
\label{eq:hmunusolved}
\frac{1}{a^2_{h,v}} \Box^{(4)} \bar{h}_{\mu \nu}^{h,v}(y=0,y_c)=
-16 \pi \, G \sum_{b=h,v} a^2_b \left[T^b_{\mu \nu} -\frac{1}{2} \eta_{\mu \nu}
T^b_{\sigma \rho} \eta^{\sigma \rho} \right],
\eeq
where $b=h,v$ indicates quantities evaluated on the hidden or visible branes,
$\Box^{(4)}=\eta^{\mu \nu} \partial_\mu \partial_\nu$,
$T$ is the energy-momentum tensor and
$G=\left(2 M_5^3 \int_{y=0}^{y=y_c} a^2(y) dy \right)^{-1}$.
The transverse traceless part of $h$ is denoted  by $\bar{h}$.

We are interested in static point sources, for which 
\beq
T^{\mu\nu}=m\, U^\mu U^\nu \delta^{(3)}(x)/\sqrt{-g^{(3)}}.
\eeq
Here $m$ is the mass of the source particle, $U^\mu=dx^\mu/d\tau$ is
its four velocity, which we take to be only in the time
direction. Therefore on $y=0$ one has $U^0=a^{-1}(0)$, while on
$y=y_c$ it is $U^0=a^{-1}(y_c)$:
\beq
\label{eq:sources}
T_{h\,00}=m_o\delta^{(3)}(x)\qquad
T_{v\,00}=m_ca^{-1}(y_c)\delta^{(3)}(x).
\eeq 
Putting a source at $\vec x=0$ on either brane we obtain
\beq
\bar{h}_{00}(x)= m_{h,v} 2 G a^3_{h,v} \frac{1}{|\vec x|}.
\eeq

For a particle motion restricted to the brane we are only interested in the
geodesic equations for the $x^\mu$ components. Since the metric is static,
the only nonzero component of the affine connection linear in the
perturbation is $\Gamma^i_{00}=-\frac{1}{2 a^2} \bar{h}_{00,i}$.
Therefore, the geodesic equation reads
\beq
\frac{d^2\vec x}{d\tau^2} =-m_{h,v} G a_{h,v} \frac{\vec x}{|\vec x|^3}
\left(\frac{d t}{d\tau}\right)^2.
\eeq
We need to express the distances in terms of the physical distance
$\vec x_{\rm phys}= a_{h,v} \vec{x}$. Finally we obtain for the physical
acceleration towards the source
\beq
\frac{d^2\vec x_{\rm phys}}{dt^2} =-m_{h,v} G a^2_{h,v}
\frac{\vec x_{\rm phys}}{|\vec x_{\rm phys}|^3},
\eeq
which implies that the effective Newton's constant measured
by brane observers are
\beq
\label{eq:GN}
G_N^{h,v}=a^2_{h,v} G.
\eeq
On the surface this result follows directly from
Eq.~(\ref{eq:hmunusolved}), but the derivation shows that many
additional warp
factors come in and conspire  to give Eq.~(\ref{eq:GN}) only when
describing the acceleration in terms of physical lengths. 

Comparing this result with the previous section we see that on the hidden
brane all interactions are governed by the same fundamental scale. On
the visible brane, brane fields have masses equal to the fundamental scale,
while the Newton's constant is suppressed compared to the fundamental scale.
Alternatively, the Planck scale inferred by an observer on the visible brane
is enhanced compared to the fundamental scale of five-dimensional gravity.

If we want this model to reproduce the observed hierarchy we need to set
the fundamental scale of five-dimensional quantum gravity to be of the order
of the weak scale. Of course, this  situation is similar to the models
with large extra dimensions~\cite{largexd}. The crucial distinction
between the RS scenario and the large extra dimensions is that the size
of the extra dimension in the RS model is only a small multiple
of the fundamental scale. Therefore, only a small fine tuning is required to
obtain the proper hierarchy (apart from tuning the brane tensions to
the bulk cosmological constant). It has already been noted in Ref.~\cite{RS}
that higher-dimensional terms in the Lagrangian are suppressed
by powers of the weak scale, so experiments in the near future
should see signals of new physics.

\section{Kaluza-Klein Modes}
\label{sec:KK}
For completeness we study the masses of bulk scalar fields. We want to
make sure that the hierarchy on the visible brane is not upset
by an emergence of a new mass scale~\cite{GW:scalar}.
Our strategy is similar to that presented in Section.~\ref{sec:rescaling}.
We analyze the full two-point function of a scalar bulk field.
We then restrict the full Green function to observations on the
branes, that is, 
with both arguments of the Green function set to either $y=0$ or $y=y_c$.
We then express the $x^\mu$ coordinates in terms of invariant distances on
the relevant brane.

We therefore turn our attention to the Green function of the
Klein-Gordon equation,
\beq
\label{eq:KGGreen}
(\Box + m^2) \Delta(x,y;x',y')=\frac{\delta^4(x-x')\delta(y-y')}{\sqrt
G}.  
\eeq
Here $\Box=\frac1{\sqrt G}\partial_A\sqrt GG^{AB}\partial_B$ is the
scalar Laplacian in the space~(\ref{eq:RSmetric}).  For simplicity we
first consider the case with $m=0$; $m\ne0$ is presented before the
end of this section. The details of the derivation are relegated to
the Appendix. As shown there
\beq
\label{eq:fullgreen}
\Delta(x,y;x',y')=\int\frac{d^4q}{(2\pi)^4}e^{iq\cdot(x-x')}
\Delta_q(y,y'),
\eeq
where
\beq
\Delta_q(y,y')=\frac{\pi k^3(zz')^2}2
\frac{[N_1(\hat qz_2)J_2(\hat qz_>)-J_1(\hat qz_2)N_2(\hat qz_>)]
[N_1(\hat qz_1)J_2(\hat qz_<)-J_1(\hat qz_1)N_2(\hat qz_<)]}%
{N_1(\hat qz_1)J_1(\hat qz_2)-J_1(\hat qz_1)N_1(\hat qz_2)}.
\eeq
Here $J_n$ and $N_n$ are  Bessel functions of the first kind and
Neumann functions, respectively,
and $\hat q\equiv\sqrt{\eta_{\mu\nu}q^\mu q^\nu}$. We have introduced the
conformal variable
\beq
\label{eq:zdefd}
z=\frac1ke^{ky},
\eeq
the brane values $z_{1,2}=z|_{y=0,y_c}$, and the notation $z_>$ and
$z_<$ to represent the larger and smaller of $z$ and $z'$,
respectively. 

The function $\Delta_q(y,y')$ has isolated poles at $\hat q=0$ and at
$\hat q=m_n>0$, where $m_n$ are solutions to 
\beq
N_1(m_nz_1)J_1(m_nz_2)-J_1(m_nz_1)N_1(m_nz_2)=0.
\eeq
It is easy to verify that,  for low excitation number $n$, $m_n\sim
a(y_c)k$. Denoting the residues of the poles by $-R_n(y,y')/2m_n$ we
have
\bea
\label{eq:Deltaexpansion}
\Delta(x,y;x',y')&=&-\sum_n\int\frac{d^4q}{(2\pi)^4}
\frac{e^{iq\cdot(x-x')}}{q^2-m_n^2}R_n(y,y')\nonumber\\
&=&\sum_n\Delta^{(4)}(x-x';m_n)R_n(y,y'),
\eea
where $\Delta^{(4)}(x-x';m_n)$ is the four dimensional Green function
for a particle of mass $m_n$. Since the residue factorizes,
$R_n(y,y')=r_n(y)r_n(y')$, the full Green function $
\Delta(x,y;x',y')$ can be obtained in a four dimensional description
by coupling sources $J(x)$ and $J'(x)$ to the linear combinations
$\sum_n r_n(y)\psi_n(x)$ and $\sum_n r_n(y')\psi_n(x)$, respectively,
where $\psi_n$ is a field of mass $m_n$. 

By the arguments of the previous section, the result of expressing
distances in terms of the invariant distances an observer on the visible
(negative tension) brane sees particles of physical mass
$m_n/a(y_c)\sim k$. Their `overlap', or wave-function on the visible
brane, is given by $r_n(y_c)$. It is only observers on the {\it hidden}
(positive tension) brane who see exponentially suppressed masses. The
calculation can be easily repeated using the rescaled
metric~(\ref{eq:RSmetricrescaled}), and the same conclusions are reached.
There is a simple physical interpretation. Hidden brane observers see
masses that have climbed up a potential well and are therefore red-shifted
compared to the visible brane precisely by the warp factor.

This discussion goes through with little modification in the case of
massive bulk scalars. Using the Fourier transform
in~(\ref{eq:fullgreen}), the solution to Eq.~(\ref{eq:KGGreen}) is
\beq
\label{eq:fullgreenmassive}
\Delta_q(y,y')=\frac{\pi k^3(zz')^2}{2}
\frac{[\tilde N_\nu(\hat qz_2)J_\nu(\hat qz_>)
-\tilde J_\nu(\hat qz_2)N_\nu(\hat qz_>)]
[\tilde N_\nu(\hat qz_1)J_\nu(\hat qz_<)
-\tilde  J_\nu (\hat qz_1)N_\nu(\hat qz_<)]}%
{\tilde N_\nu(\hat qz_1)\tilde J_\nu(\hat qz_2)
-\tilde J_\nu (\hat qz_1)\tilde N_\nu (\hat qz_2)}.
\eeq
where $z$ is given by Eq.~(\ref{eq:zdefd}), $\nu=\sqrt{4+m^2/k^2}$,
and we have introduced the shorthand
\beq
\tilde Z_\nu(z) = (1-\frac\nu2)Z_\nu(z)+\frac{z}2Z_{\nu-1}(z),
\eeq
for $Z$ a Bessel function. The function $\Delta_q(y,y')$ is regular
at $q=0$. However it diverges at $q=0$ as $m\to0$. Therefore, there is
no massless particle in the Kaluza-Klein spectrum except for
$m=0$. The spectrum is determined by the zeroes of the denominator,
\beq
{\tilde N_\nu(m_nz_1)\tilde J_\nu(m_nz_2)
-\tilde J_\nu (m_nz_1)\tilde N_\nu (m_nz_2)}=0.
\eeq
This is precisely the same equation as found by Goldberger and Wise by
means of a different method\cite{GW:scalar}, namely, direct
diagonalization of the action integral. For small $m$, the low
excitation number spectrum has $m_n\sim a(y_c)k$. As above, the full
Green function can be written as a sum over poles,
Eq.~(\ref{eq:Deltaexpansion}), where the residues factorize.
The physical picture is the same as in the massless case.
For $m$ of the order the fundamental scale, the lowest mass state that
an observer on the visible brane  measures has mass of order $m$, 
while an observer on the hidden brane measures a mass of order $a(y_c) m$.

\section{Summary}
\label{sec:summary}
We have discussed a covariant procedure for determining masses in brane
scenarios. The results are independent of coordinate rescaling.  We
analyzed the situation from the point of view of brane observers who
measure distances along branes using, as they must, the induced
metric. Rather than absorbing the metric into redefinitions of matter
fields we soaked up the warp factors by expressing all coordinate
quantities in terms of invariant distances.
 
Applying the covariant approach to the Randall-Sundrum solution of the
hierarchy problem one finds that the observed masses for brane fields
coincide with their Lagrangian values. The hierarchy is realized due
to a suppression of gravitational interactions on the visible
brane. For observers on the visible (negative tension) brane the
effective value of Newton's constant is suppressed by the warp factor
$a(y)$. This suppression is a result of the overlap of the graviton
wave function with the brane fields\cite{RS}. It is important for
maintaining the hierarchy that the long-range forces are mediated only
by the four-dimensional graviton. In the absence of a mechanism for
radius stabilization the radion field would provide the dominant
contribution to the static gravitational force on the visible brane.

In the covariant approach it is natural to always compare masses with
the fundamental scale of the underlying five-dimensional theory $M_5$.
All bulk fields, not only the graviton, look differently to
observers on different branes. The towers of Kaluza-Klein states for
bulk fields are observed to have different spacings on different
branes. Masses measured on different branes are related to each other
by the ratio of the corresponding warp factors.  This is a result of
climbing or falling into the gravitational potential.

The RS model solves the hierarchy problem by naturally assuming that
all mass parameters in the underlying Lagrangian, including the brane
Lagrangians, must be of order $M_5$. Since the observed masses for
brane fields coincide with their Lagrangian values, $M_5$ must be
around the weak scale to ensure the weak scale vacuum expectation
value for the Higgs field.

\bigskip
{\it Acknowledgments} 
We would like to thank Ken Intriligator and Lisa Randall for discussions.
We would like to thank Csaba Cs\'aki, Ira Rothstein, and  Raman Sundrum
for their comments on the manuscript.
While completing this paper we become aware of Ref.~\cite{Ozeki}
where ideas somewhat similar to ours were pursued.
The work of B.G.\ is supported by the U.S. Department of
Energy under contract No.\ DOE-FG03-97ER40546, the work of D.N.\ 
under contract No.\ DE-FG02-90ER40542, and the work of W.S. under cooperative
research agreement DE-FC02-94ER40818.

\appendix
\section{Bulk Scalar Green Functions}
For completeness we derive the Green function of a massless and a massive
bulk scalar field. We then comment on the results of
Ref.~\cite{GW:scalar}.
We start with the massless case because it is simpler. Our derivation
follows the derivation of Ref.~\cite{GKR}. The only
difference from  Ref.~\cite{GKR} is in the boundary
conditions since we are analyzing a two-brane scenario.

In the weak coupling limit the scalar field does not affect the metric.
Thus, we solve the equation
\begin{equation}
\label{Greendef}
  \left( \frac{1}{\sqrt{G}} \partial_M \sqrt{G} G^{MN} \partial_N \right) 
  \Delta(x,z,x',z') =\frac{\delta^4(x-x') \delta(z-z')}{\sqrt{G}}
\end{equation}
with fixed background metric
\begin{equation}
ds^2=  G_{MN}dx^Mdx^N=\frac{1}{(kz)^2} (\eta_{\mu\nu}dx^\mu dx^\nu-dz^2).
\end{equation}
Here we use conformaly flat coordinates with $z=\frac{1}{k}
e^{ky}$. For the calculation in the rescaled
metric~(\ref{eq:RSmetricrescaled}) simply take $z=\frac{1}{k}
e^{k(y-y_c)}$.
With this choice of coordinates the 5-dimensional Laplacian has the
form
\begin{equation}
  \frac{1}{\sqrt{G}} \partial_M \sqrt{G} G^{MN} \partial_N =
  (kz)^2 \eta^{\mu\nu} \partial_\mu \partial_\nu -
  (kz)^5 \partial_z \frac{1}{(kz)^3} \partial_z.
\end{equation}
We first Fourier transform the Green function 
\begin{equation} 
  \Delta(x,z,x',z')=
  \int \frac{d^4 q}{(2 \pi)^4} e^{i q (x-x')} \Delta_q(z,z').
\end{equation}
In the momentum space
Eq.~(\ref{Greendef}) takes the form
\begin{equation}
  \frac{1}{(kz)^3}\left( -q^2 - \partial_z^2 -\frac{3}{z} \partial_z \right)
  \Delta_q(z,z') =  \delta(z-z').
\end{equation}
where $q^2=\eta^{\mu\nu} q_\mu q_\nu$. 
Introducing $\hat{\Delta}_q =(\frac{1}{k^2zz'})^2 \Delta_q$,
like in Ref.~\cite{GKR}, the
above equation becomes the Bessel equation
\begin{equation}
\label{Besselm0}
  (z^2 \partial_z^2 + z \partial_z + q^2 z^2 - 4) \hat{\Delta}_q
  =-  \frac{z}{k} \delta(z-z').
\end{equation}

The standard method for solving this kind of equation is to first find
solutions to the homogeneous equation in two regions $z<z'$ and $z>z'$.
Let us call these solutions $\hat{\Delta}_<$ and  $\hat{\Delta}_>$,
respectively.
Then the full equation is solved by requiring that the discontinuity of
the first derivative at $z=z'$ reproduces the delta function.
The second order Bessel functions $J_2(q z)$ and $N_2(q z)$ are the
linearly independent solutions to the homogeneous part of Eq.~(\ref{Besselm0}).

We impose the Neumann boundary conditions at $z=R_<$ and $z=R_>$ on the
original Green function $\partial_z \Delta |_{z=R_<,R_>}=0$.
$R_<$ corresponds to the hidden brane,  $z=R_>$ to the visible one.
We obtain
\begin{eqnarray}
 \hat{\Delta}_< &=& A_<(z')
            \left[N_1(q R_<) J_2(q z) - J_1(q R_<)N_2(q z) \right], \\
 \hat{\Delta}_> &=& A_>(z')
            \left[N_1(q R_>) J_2(q z) - J_1(q R_>)N_2(q z) \right].
\end{eqnarray}
We could have imposed the boundary conditions taking the derivatives
with respect to $z'$ instead of $z$ and correspondingly exchanged the
subscripts $<$ with $>$. Therefore, by symmetry we can write
\begin{eqnarray}
 \hat{\Delta}_< &=& C_<
   \left[N_1(q R_>) J_2(q z') - J_1(q R_>)N_2(q z') \right]
   \left[N_1(q R_<) J_2(q z) - J_1(q R_<)N_2(q z) \right], \\
 \hat{\Delta}_> &=& C_> 
   \left[N_1(q R_<) J_2(q z') - J_1(q R_<)N_2(q z') \right]
   \left[N_1(q R_>) J_2(q z) - J_1(q R_>)N_2(q z) \right],
\end{eqnarray}
where $C_<$ and $C_>$ are constants.

The solution has to be continuous at $z=z'$, so
\begin{equation}
  C_<=C_>\equiv C.
\end{equation} 
Moreover, the first derivative must be discontinuous
\begin{equation}
 kz \partial_z (\Delta_> -  \Delta_<)|_{z=z'} =1,
\end{equation}
which implies that
\begin{eqnarray}
  1&=& k q z C \Big[ \left[ N_1(q R_<) J_2(q z) - J_1(q R_<)N_2(q z) \right]
   \left[ N_1(q R_>) J'_2(q z) - J_1(q R_>)Y'_2(q z) \right] \\
   && - \left[N_1(q R_>) J_2(q z) - J_1(q R_>)N_2(q z) \right]
   \left[N_1(q R_<) J'_2(q z) - J_1(q R_<)Y'_2(q z) \right] \Big].
\end{eqnarray}
Using the identities $z Z'_\nu= z Z_{\nu-1} - \nu Z_{\nu}$, where 
$Z$ stands for either $J_\nu$ or $N_\nu$ and
$N_1(z) J_2(z) - J_1(z) N_2(z) =\frac{2}{\pi z}$ we obtain
\begin{equation}
 \frac{1}{C} =\frac{2k}{\pi} 
 \left[ J_1(q R_>) N_1(q R_<) - N_1(q R_>) J_1(q R_<) \right] .
\end{equation}

Solving the massive case is now simple. We need to Fourier transform
in the $x$ variables and perform the redefinition from $\Delta_q$
to $\hat{\Delta}_q$. The only modification of Eq.~(\ref{Besselm0})
is the mass term
\begin{equation}
\label{Besselmnot0}
  (z^2 \partial_z^2 + z \partial_z + q^2 z^2 - 4-m^2) \hat{\Delta}_q
  =-  \frac{z}{k} \delta(z-z').
\end{equation}
The solutions to the homogeneous part are $J_\zeta(q z)$ and 
$N_\zeta(q z)$, where $\zeta=\sqrt{4 + m^2}$.

Again, we impose the Neumann boundary conditions at $z=R_<$ and
$z=R_>$. When taking derivatives of $J_\zeta$ and $N_\zeta$
we encounter the linear combinations
\begin{eqnarray}
  \tilde{J}_\zeta(x) &\equiv& (1 -\frac{\zeta}{2}) J_\zeta(x) + 
        \frac{x}{2} J_{\zeta-1} (x), \\
  \tilde{N}_\zeta(x) &\equiv& (1 -\frac{\zeta}{2}) N_\zeta(x) + 
        \frac{x}{2} N_{\zeta-1} (x).
\end{eqnarray}
In terms of these newly defined variables equations for the massive
case look similar to the massless ones. Imposing the boundary
conditions gives
\begin{eqnarray}
 \hat{\Delta}_< &=& C_<
   \left[\tilde{N}_\zeta(q R_>) J_\zeta(q z') 
         - \tilde{J}_\zeta(q R_>)N_\zeta(q z') \right]
   \left[\tilde{N}_\zeta(q R_<) J_\zeta(q z) 
         - \tilde{J}_\zeta(q R_<) N_\zeta(q z) \right], \\
 \hat{\Delta}_> &=& C_> 
   \left[\tilde{N}_\zeta(q R_<) J_\zeta(q z') 
         - \tilde{J}_\zeta(q R_<)N_\zeta(q z') \right]
   \left[\tilde{N}_\zeta(q R_>) J_\zeta(q z) 
         - \tilde{J}_\zeta(q R_>)N_\zeta(q z) \right].
\end{eqnarray}
 
In complete analogy to the massless case, continuity at $z=z'$
requires $C\equiv  C_<=C_> $. The difference in the first
derivatives fixes $C$ to be
\begin{equation}
\label{denominatormnot0}
 \frac{1}{C} =\frac{2k}{\pi} 
 \left[ \tilde{J}_\zeta(q R_>) \tilde{N}_\zeta(q R_<) 
       - \tilde{N}_\zeta(q R_>) \tilde{J}_\zeta(q R_<) \right] .
\end{equation}

As we discussed in Sec.~\ref{sec:KK}  the poles of Green functions,
or zeros of Eq.~(\ref{denominatormnot0}), correspond to physical
masses after proper rescaling. We want to comment of the results 
of Ref.~\cite{GW:scalar}. Goldberger and Wise consider a free
scalar field bulk action
\begin{eqnarray}
\label{eq:GWaction}
S&=&{1\over 2}\int d^4 x dy \sqrt{G} \left(G^{AB}\partial_A \Phi
\partial_B \Phi - m^2 \Phi^2\right)\nonumber\\
&=&{1\over 2}\int d^4 x dy
\left(e^{-2k|y|}\eta^{\mu\nu}\partial_\mu\Phi \partial_\nu \Phi +
\Phi\partial_y\left(e^{-4k|y|}\partial_y\Phi\right)
-m^2 e^{-4k|y|}\Phi^2\right), 
\end{eqnarray}
The Kaluza-Klein decomposition is in terms of the modes
\beq
\Phi(x,y)=\sum_n \psi_n(x) \xi_n(y),
\eeq
which satisfy
\beq
\int_{0}^{y_c} dy e^{-2k|y|} \xi_n(y) \xi_m(y)=\delta_{nm}
\eeq
and
\beq
\label{eq:GWeigen}
-\frac{d}{dy}\left(e^{-4k|y|}\frac{d\xi_n}{dy}\right)
+m^2 e^{-4k|y|}\xi_n =m_n^2 e^{-2k|y|} \xi_n.
\eeq
Inserting this in Eq.~(\ref{eq:GWaction}) gives
\beq
\label{eq:effaction}
S={1\over 2}\sum_n \int d^4 x \left[\eta^{\mu\nu}\partial_\mu \psi_n
\partial_\nu \psi_n - m_n^2 \psi_n^2\right]. 
\eeq
Solving Eq.~(\ref{eq:GWeigen}) for $m_n$ gives precisely our
Eq.~(\ref{denominatormnot0}). The explanation is that
any procedure for diagonalizing the action should lead to
the same eigenvalues. However, $m_n$ are not coordinate-invariant
quantities. In our covariant interpretation, $m_n$ are the physical masses
on the hidden brane, while on the visible brane $m_n/a(y_c)$ are the
measured masses.


\end{document}